\documentclass[a4paper, amsfonts, amssymb, amsmath, reprint, showkeys, nofootinbib, twoside, superscriptaddress]{revtex4-1}
\usepackage[english]{babel}
\usepackage[utf8]{inputenc}
\usepackage{times} 
\usepackage[colorinlistoftodos, color=green!40, prependcaption]{todonotes}
\usepackage{graphicx}
\usepackage{xfrac}
\usepackage{bm}
\usepackage{float}
\usepackage{newfloat,algcompatible}
\graphicspath{{Images/}}
\usepackage[pdftex, pdftitle={Article}, pdfauthor={Author}]{hyperref} 
\usepackage{url}
\usepackage{hyperref}
\hypersetup{
    colorlinks=true,
    linkcolor=blue,
    filecolor=magenta,      
    urlcolor=blue,
    citecolor=blue,
    pdftitle={Overleaf Example},
    pdfpagemode=FullScreen,
    }

\DeclareFloatingEnvironment[
    fileext=loa,
    listname=List of Algorithms,
    name=ALGORITHM,
    placement=tbhp,
]{algorithm}
\DeclareMathOperator{\sinc}{sinc}

\begin{document}

\title{Differentiable master equation solver for quantum device characterisation}

\author{D.L.~Craig}
\affiliation{Department of Materials, University of Oxford, Parks Road, Oxford OX1 3PH, United Kingdom}

\author{N.~Ares}
\affiliation{Department of Engineering Science, University of Oxford, Parks Road, Oxford OX1 3PJ, United Kingdom}

\author{E.M.~Gauger$^*$}
\affiliation{SUPA, Institute of Photonics and Quantum Sciences, Heriot-Watt University, Edinburgh EH14 4AS, United Kingdom}

\begin{abstract}
Differentiable models of physical systems provide a powerful platform for gradient-based algorithms, with particular impact on parameter estimation and optimal control. Quantum systems present a particular challenge for such characterisation and control, owing to their inherently stochastic nature and sensitivity to environmental parameters. To address this challenge, we present a versatile differentiable quantum master equation solver, and incorporate this solver into a framework for device characterisation. Our approach utilises gradient-based optimisation and Bayesian inference to provide estimates and uncertainties in quantum device parameters. To showcase our approach, we consider steady state charge transport through electrostatically defined quantum dots. Using simulated data, we demonstrate efficient estimation of parameters for a single quantum dot, and model selection as well as the capability of our solver to compute time evolution for a double quantum dot system. Our differentiable solver stands to widen the impact of physics-aware machine learning algorithms on quantum devices for characterisation and control.
\end{abstract}

\maketitle

\def\thefootnote{$*$}\footnotetext{e.gauger@hw.ac.uk} 

\section{Introduction}
\label{sec:introduction}

The challenge of characterising quantum devices has inspired a myriad of approaches, with the goal of furthering the understanding and application of quantum technologies \cite{gebhart2023learning}. The simplest approach to characterising aspects of a quantum system is phenomenological fitting, which can yield useful parameter values with minimal resources and carefully selected measurements. Phenomenological fitting has the advantage of being readily applied to suitable data, but neglects detailed structure of the model describing a physical system and often fails when closed-form analytic solutions are unknown. More complex methods such as quantum state tomography \cite{leonhardt1995quantum} and quantum process tomography \cite{mohseni2008quantum} facilitate full characterisation of the states prepared in a device or how they are changed by device operation, but often require large amounts of data as well as access to observables providing a full tomographic basis set. Several approaches have been developed to improve the efficiency of these tomographic techniques, or to extend their descriptive power, by having appropriate models incorporated into the characterisation of both closed \cite{cole2005identifying,devitt2006scheme,zhang2014quantum} and open quantum systems \cite{jorgensen2019exploiting,white2022non}. 

Deep learning methods have recently opened many promising avenues for characterising quantum systems. For instance, efficient tomographic methods have been developed using deep learning models as a variational ansatz \cite{torlai2018neural}, and deep learning methods have also been applied to characterisation of open quantum system dynamics and estimating physical parameters describing the system and environment \cite{banchi2018modelling,hartmann2019neural,flurin2020using, barr2023spectral}. Learning an abstracted generator of quantum dynamics, such as in the case of deep learning, may lead to accurate prediction of a quantum system's evolution but often fails to provide direct insight into physical parameters describing the system. Capitalising on prior knowledge about the dominant physical interactions allows for more explicit and efficient characterisation of quantum systems than abstracted tomographic models and with greater flexibility than phenomenological fitting \cite{kurchin2024using}. In such a physically motivated paradigm, parameter values for a physical model could be inferred from available measurements, and other aspects of the system could be probed by inserting inferred values into the same model. 

The success of modern deep learning methods is underpinned by differentiable programming. This programming paradigm facilitates the automatic differentiation required for training deep neural networks with many parameters, where numerical and symbolic derivatives become severely impractical \cite{rumelhart1986learning,baydin2018automatic}. Beyond applications to deep learning, differentiable programming has recently been applied to physical models in the context of physics-aware machine learning \cite{craig2024bridging,pachalieva2022physics,wright2022deep}, efficient computation of gradients with respect to complex physical models \cite{liao2019differentiable,Hu2020DiffTaichi:}, optimisation of non-equilibrium steady states \cite{vargas2021fully}, and characterisation of non-Markovian dynamics \cite{krastanov2020unboxing}. Deep learning methods have recently been combined with differentiable models of quantum systems to enhance quantum device characterisation \cite{genois2021quantum, youssry2024experimental}. There have also been recent advances in using differentiable models for simulations of quantum systems with a focus on optimal control \cite{schafer2020differentiable,coopmans2021protocol,wittler2021integrated,khait2022optimal,coopmans2022optimal, sauvage2022optimal}, a requirement for the future of pulse engineering \cite{liang2022variational}.

\begin{figure*}[ht]

    \centering
    \includegraphics[width=0.96\textwidth]{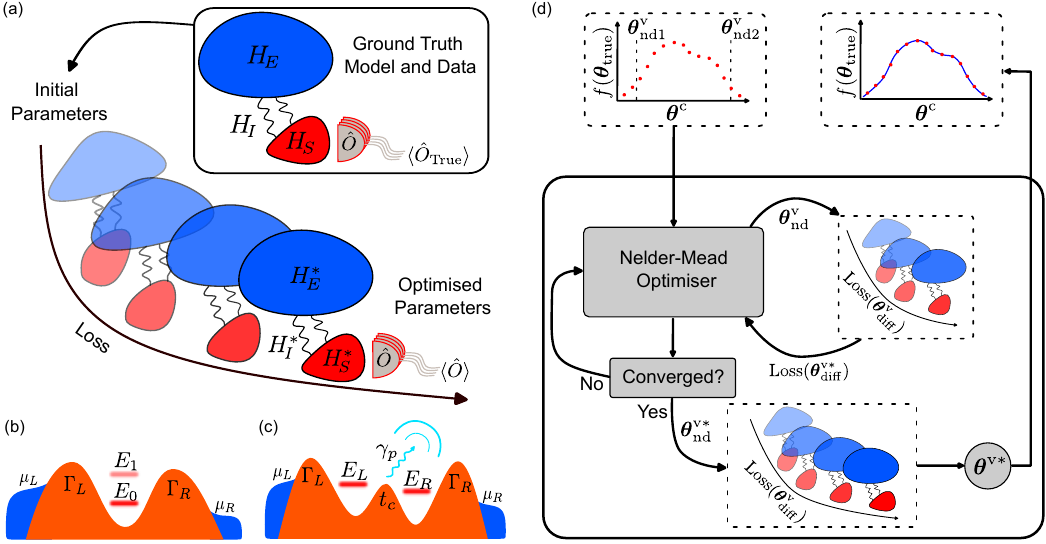}
        
\caption{(a) A schematic of the process of estimating parameters of an open quantum system using gradient descent. Here, $H_\mathrm{S}$, $H_\mathrm{E}$ are the system and environment Hamiltonians, respectively, and $H_\mathrm{I}$ is the system-environment interaction Hamiltonian. Expectation values of an observable $\langle \hat{O} \rangle$ are generated from a ground truth model, which would be experimental data in practical applications, and a differentiable model is used to fit these values by gradient descent of a loss function. (b) Representation of the first case study of a single quantum dot with an excited state coupled to left and right leads. Here, $\mu_{\mathrm{L}(\mathrm{R})}$ is the chemical potential of the left(right) lead, $\Gamma_{\mathrm{L}(\mathrm{R})}$ is the tunnel rate between the dot states and the left(right) lead, and $E_{0(1)}$ is the electrochemical potential of the dot in the ground(excited) state. (c) Representation of the second case study of a double quantum dot coupled to left and right leads, and a phonon bath. In each case study the observable used to inform parameter estimation is the steady state current from left to right lead. Here $E_{\mathrm{L}(\mathrm{R})}$ is the chemical potential of the dot with a charge on the left(right) dot, $t_c$ is the tunnel coupling between the dots, and $\gamma_p$ is the coupling strength to the phonon bath. (d) An illustrative schematic of our approach for optimising differentiable and non-differentiable parameters when fitting to data. Here, we denote non-differentiable parameters as $\boldsymbol{\theta}^\mathrm{v}_\mathrm{nd}$, differentiable parameters as $\boldsymbol{\theta}^\mathrm{v}_\mathrm{diff}$, and optimised parameters are indicated with an asterisk.}
\label{fig:algorithm_sketch}
\end{figure*}

In this article we apply the concepts of differentiable programming to develop a versatile, fast and differentiable quantum master equation solver. We apply our solver to electrostatically defined quantum dots as they offer a multi-faceted platform for realising quantum systems for which accurate parameter estimation would benefit experiments in quantum computation \cite{hanson2007spins,burkard2023semiconductor,chatterjee2021semiconductor}, quantum thermodynamics \cite{josefsson2018quantum}, and environment-assisted transport \cite{granger2012quantum,hartke2018microwave}. The interaction of discrete charge states with fermionic leads and vibrational modes requires an open quantum systems description, for which weak-coupling master equations can be a suitable choice in many experimental device regimes \cite{stace2005population,hartke2018microwave,hofmann2020phonon}. Measurements of (steady state) charge transport in single and double quantum dot systems are often used to estimate characteristics of the system, such as temperature \cite{beenakker1991theory,prance2012single,borsoi2023shared}, energy spectra, \cite{foxman1993effects, kouwenhoven1997electron}, and electron-phonon coupling mechanisms \cite{hartke2018microwave,hofmann2020phonon, sowa2017environment}. However, characterisation of electrostatically defined quantum dot devices remains challenging, not least due to the fact that control via voltages applied to gate electrodes introduces an abstraction between experimental control and system parameters \cite{ares2021machine, zwolak2023colloquium, gebhart2023learning}. Further to this abstracted control, disorder frustrates device characterisation by causing variation in the control parameters for devices of the same design \cite{craig2024bridging, chatzikyriakou2022unveiling, percebois2021deep, seidler2023tailoring}.

To address such demanding situations, our approach combines a differentiable quantum master equation solver with gradient-based optimisation protocols and Bayesian inference, and is designed to handle a mixture of differentiable and non-differentiable parameters to facilitate a range of applications. In addition to gradient-based optimisation of parameters, Markov chain Monte Carlo methods facilitate estimation of posterior distributions from which parameter uncertainties can be obtained. 

To demonstrate the power and broad applicability of our approach, we have designed two case studies with different levels of complexity. The features probed in our case studies are not suitable for phenomenological fitting, and each case study is suitably complex that typical analytic solutions for quantum dots are not applicable. The first case study of transport through a single quantum dot with an orbital excited state is designed to highlight our approach's capabilities in handling differentiable and non-differentiable parameters. The second case study of transport through a double quantum dot coupled to a phonon bath is designed to explore characterisation of the environment \cite{gullans2018probing,hofmann2020phonon,barr2023spectral}. In this second case study, we showcase the ability of our algorithm to discriminate between and correctly select from a set of models featuring different phonon spectral densities relating to different device materials and geometries. To quantitatively assess the performance of our approach, these case studies utilise simulated data. A sketch of the differentiable fitting process, which is central to our approach, is shown in Figure~\ref{fig:algorithm_sketch}(a) along with both case study scenarios in (b) and (c), respectively, and an outline of our approach in (d).

This article is structured as follows:
First, we discuss our implementation of a differentiable quantum master equation solver in Section~\ref{sec:differentiable}, before discussing the Bayesian formulation of our parameter estimation algorithm in Section~\ref{sec:bayesian} and optimisation of non-differentiable parameters in Section~\ref{sec:non_differentiable}. We apply our approach to case studies of single and double quantum dots in Section~\ref{sec:case_study_single} and Section~\ref{sec:case_study_double} respectively. Finally, in Section~\ref{sec:time_evolution} we demonstrate the capability of our model to calculate time evolution of a double quantum dot system.

\section{Differentiable Master Equation Solver}
\label{sec:differentiable}

The focal point of our quantum device characterisation is a differentiable quantum master equation solver, which is implemented using TensorFlow \cite{abadi2016tensorflow}. Quantum master equations are a method of determining the time evolution of the system density matrix $\rho$ of an open quantum system \cite{breuer2002theory}. In a weak-coupling setting this evolution is governed by the Liouvillian superoperator, $\mathcal{L}$, which contains details of both the system Hamiltonian and the system-environment interaction, such that $\dot\rho(t)=\mathcal{L}\rho(t)$. A simple manifestation of $\mathcal{L}$ is given by a Lindblad master equation, which reads
\begin{equation}
    \label{lindblad}
    \dot\rho(t) = -i[H_S, \rho(t)] + \sum_i \Gamma_i \mathcal{D}[A_i]\rho(t),
\end{equation}
where $H_S$ is the system Hamiltonian, $\Gamma_i$ is the dissipation rate associated with the operator $A_i$ which describes environment induced transitions in the system, and $\mathcal{D}[A]\rho = A\rho A^\dag - \sfrac{1}{2}\{A^\dag A,\rho\}$ is the Lindblad dissipator acting on the density matrix. The dissipation rates are often dependent on system Hamiltonian parameters defining the energy difference between states involved in the environment induced transitions. 

Numerically solving such a quantum master equation using differentiable programming facilitates evaluation of the gradient of an observable with respect to the master equation parameters. To compute steady states, we solve the set of linear equations resulting from Eq.~(\ref{lindblad}) under the condition $\dot\rho(t)=0$, and for time evolution we use ODE solvers as discussed in Section~\ref{sec:time_evolution}. A differentiable model allows gradients to be obtained with similar computation time to forward evaluation of the model using automatic differentiation, without the need for inefficient finite-element methods. Solving the steady state directly from the Liouvillian is critical to the speed of our solver by avoiding the many operations and storing of gradients required for a long-time integration of an ODE. Our implementation can deal with arbitrary Liouvillian superoperators, and is thus applicable to a range of physical systems. 

Using TensorFlow allows direct use of in-built gradient-based optimisation methods such as Adam \cite{kingma2014adam}, and Bayesian inference methods such as Hamiltonian Monte Carlo (HMC) \cite{neal2011mcmc} which can be implemented using TensorFlow Probability \cite{dillon2017tensorflow}. Several differentiable programming libraries, including TensorFlow, have the advantage of allowing vectorised (i.e. batched) inputs and computation performed on a graph which leads to significant speed-up when compared to the same computation performed using a standard library for modelling quantum systems, such as QuTiP \cite{johansson2012qutip}. Vectorised inputs refer to having the input to the differentiable model be an $n_\mathrm{b}\times n_\theta$ array for $n_\theta$ parameters and a batch of $n_\mathrm{b}$ different sets of parameter values, so that the output will be a $n_\mathrm{b}\times n_\mathrm{out}$ array, where $n_\mathrm{out}$ is the number of outputs from the model. An example demonstrating the accuracy and speed of our solver is outlined in Appendix~\ref{app:benchmarking}. 

Related works have explored differentiable master equation solvers for time dynamics \cite{krastanov2020unboxing}, and incorporated differentiable ODE solvers for learning dynamics of quantum systems with recurrent neural networks \cite{genois2021quantum}. The effectiveness of differentiable models of quantum systems enhanced by machine learning is further demonstrated in \cite{youssry2024experimental} in the context of photonic devices. Our approach takes a different direction to these works by considering steady state solutions, and implementing a data efficient characterisation algorithm which also handles non-differentiable parameters. Other related work has demonstrated differentiable computation of steady state solutions using JAX \cite{jax2018github} for quantum heat transfer models \cite{vargas2021fully}.

\section{Bayesian Estimation of Differentiable Parameters}
\label{sec:bayesian}

Our approach aims to fit model predictions to data by estimating values for parameters in quantum master equations. We consider case studies of transport through a single quantum dot with an orbital excited state, and through a double quantum dot coupled to a phonon bath. For both the single and double quantum dot case studies, the source-drain bias $V_\mathrm{bias}$ and a single 1D current scan are used to inform the parameter estimation. Not all parameters in a model may be differentiable as discussed below, and others may have minimal impact on a loss function so gradients become ineffective. We shall present our approach for dealing with such parameters in general terms before presenting results for each case study.

We denote the $n_\mathrm{c}$ parameters which are controlled to produce a batch of $n_\mathrm{b}$ data points as $\boldsymbol{\theta}^\mathrm{c}$, and the remaining $n_\mathrm{v}$ parameters under consideration for estimation as $\boldsymbol{\theta}^\mathrm{v}$. The ground truth data is then $D\! =\! \{d_j\vert j=1\ldots n_\mathrm{b}\}$ where $d_j$ is an individual data point, and the total set of $n_\theta = n_\mathrm{c} + n_\mathrm{v}$ parameters is $\boldsymbol{\theta} = [\boldsymbol{\theta}^\mathrm{c}, \boldsymbol{\theta}^\mathrm{v}]$. We assume that the ground truth data $D$ is described by a function of the true parameters $f(\boldsymbol{\theta}_\mathrm{true})$ with Gaussian noise of standard deviation $\sigma$ such that each point in $D$ is $d_j\! \sim\! \mathcal{N}(f(\boldsymbol{\theta}_\mathrm{true}),\sigma)$. Under this assumption we formulate the likelihood of the data given a set of parameters $\boldsymbol{\theta}^\mathrm{v}$ used to estimate the function $\hat{d} = f(\boldsymbol{\theta}^\mathrm{v})$ over the domain of $\boldsymbol{\theta}^\mathrm{c}$ as
\begin{equation}
    P(D\vert\boldsymbol{\theta}^\mathrm{v}) \propto \exp{\Bigg(-\frac{\sum_{j=1}^{n_\mathrm{b}}(\hat{d}_j-d_j)^2}{\sigma^2}\Bigg)}.
\end{equation}
This likelihood can be used in Bayes' formula $P(\boldsymbol{\theta}^\mathrm{v}\vert D) \propto P(D\vert\boldsymbol{\theta}^\mathrm{v})P(\boldsymbol{\theta}^\mathrm{v})$ to estimate the posterior values for parameters in $\boldsymbol{\theta}^\mathrm{v}$. The prior of each parameter in $\boldsymbol{\theta}^\mathrm{v}$ can be chosen independently such that $P(\boldsymbol{\theta}^\mathrm{v}) = \prod_{i=1}^{n_\mathrm{v}}P(\theta^\mathrm{v}_i)$. The set of parameters which produce the mode of the posterior distribution is the maximum a posteriori (MAP) estimate, and for uniform priors this becomes maximum likelihood estimation (MLE). Evaluation of the negative log probability of the posterior distribution can be used as a differentiable loss function for the gradient descent portion of the parameter fitting process discussed in the previous section. The parameters which minimise this loss correspond to the MAP estimate. 

Further to this, a Bayesian approach allows the use of Markov-chain Monte Carlo (MCMC) methods to estimate the posterior distribution for each parameter. We specifically consider Hamiltonian Monte Carlo (HMC) implemented in TensorFlow Probability \cite{dillon2017tensorflow} to generate a set of $n_s$ posterior samples for the parameter values $S=\{\boldsymbol{\theta}^\mathrm{v}_j \vert j=1\ldots n_s\}$, using the MAP estimate as a seed to reduce burn-in. These posterior samples allow for an estimate of uncertainty and correlations in the parameter values. The posterior distribution can also be normalised and used to update the prior distribution for use with further data.

\section{Optimising Non-Differentiable Parameters}
\label{sec:non_differentiable}

The output of a model is a function of the parameters, $f(\boldsymbol{\theta})$, which may not be differentiable with respect to certain elements of $\boldsymbol{\theta}$. In the case of current traces in electrostatically defined quantum dots, such parameters are the definition of the bias window in $\boldsymbol{\theta}^\mathrm{c}$ due to the abstraction between gate voltages and the energy of the dot electrochemical potential relative to the lead chemical potentials. In other situations, $f(\boldsymbol{\theta})$ may not have well-defined gradients across the whole parameter space, which renders gradient descent methods impractical.

To deal with a mixture of differentiable and non-differentiable parameters, we design an optimisation procedure which takes advantage of the fast differentiable model. The idea is to find the non-differentiable parameters for which the remaining differentiable parameters provide the best fit. A Nelder-Mead optimisation routine \cite{nelder1965simplex} finds appropriate values for the non-differentiable parameters, where the function being optimised is a fit of the model to ground truth data using the remaining differentiable parameters. This fit first involves a log-spaced grid search of the differentiable elements of $\boldsymbol{\theta}^\mathrm{v}$ where the evaluation of $f(\boldsymbol{\theta})$ with the lowest loss with respect to the ground truth data is selected as a starting point for a short gradient descent optimisation using the Adam optimiser. The evaluation of the function being optimised by the Nelder-Mead routine is simply the final loss after this gradient descent. Finally, using the optimal non-differentiable parameters, conducting a longer gradient descent optimisation of the differentiable parameters provides an estimate of all parameters in $\boldsymbol{\theta}$. A schematic of this optimisation procedure is shown in Figure~\ref{fig:algorithm_sketch}(d), and discussion of its scaling can be found in Appendix~\ref{app:algorithm_scaling}.

The model fitting algorithm discussed above is general to the situation when some non-differentiable parameters must be optimised before optimising differentiable parameters. For our case studies, the non-differentiable parameters define the energy scale of the measurement axis, which in turn defines how parameters such as tunnel rates change the profile of the fit. This is a result of dot electrochemical potentials being controlled by gate voltages in experiments, which do not have well defined values relative to the bias window. In other experimental settings, it may be the case that the measurement axis is easily quantifiable (e.g.~the frequency of an oscillating signal) and so the order in which differentiable and non-differentiable parameters are optimised becomes less important.

\section{Case Study: Single Quantum Dot with Orbital Excited State}
\label{sec:case_study_single}
\subsection*{Model}

We model a single quantum dot (SQD) with an orbital excited state as shown in Figure~\ref{fig:algorithm_sketch}(b) such that the set of charge states correspond to a maximum of one excess charge on the dot. This restriction is motivated by the large charging energy of the quantum dot compared to the bias window and the energy splitting between the ground and first orbital excited state. The available charge states are $\vert 0 \rangle$, $\vert G \rangle$, and $\vert E \rangle$, where $\vert 0 \rangle$ has an integer $M$ charges on the dot, $\vert G \rangle$ and $\vert E \rangle$ both have $M+1$ charges on the dot. The single excess charge exists in the ground state for $\vert G \rangle$, and an orbital excited state for $\vert E \rangle$. The Hamiltonian describing the coupling between these charge states and the leads is given by
\begin{align}
    H &= H_0 + H_l + H_{el}, \\
    H_0 &= E_0 \vert G \rangle \langle G \vert + \big( E_0 + \delta \big) \vert E \rangle \langle E\vert, \\
    H_l &= \sum_k (\epsilon_k - \mu_l)d_{kl}^\dag d_{kl} + (\epsilon_k - \mu_r)d_{kr}^\dag d_{kr}, \\
    H_{el} &= \sum_k \sum_{j=g,e} (t_{l} d_{kl}^\dag c_j + t_{r}d_{kr}^\dag c_j + \mathrm{H.c.}), 
\end{align}
where $E_0$ is the energy of the ground state, and $\delta$ is the energy splitting between the ground and orbital excited state. $H_l$ describes the leads with $\mu_{l(r)}$ being the chemical potential of the left(right) leads, and $d_{kl(r)}^\dag(d_{kl(r)})$ are the fermionic creation(annihilation) operators for the left(right) lead for the mode of energy $\epsilon_k$. $H_{el}$ describes the coupling between the dot states and the leads where $c_{g(e)}$ is the fermionic annihilation operator acting on the ground(excited) dot, $t_{l(r)}$ are the tunnel couplings between the leads and the dot states, and H.c. denotes the Hermitian conjugate. For simplicity, we assume the same tunnel rates to the leads for both the ground and excited charge state. 

\begin{figure}[ht]

    \centering
    \includegraphics[width=0.48\textwidth]{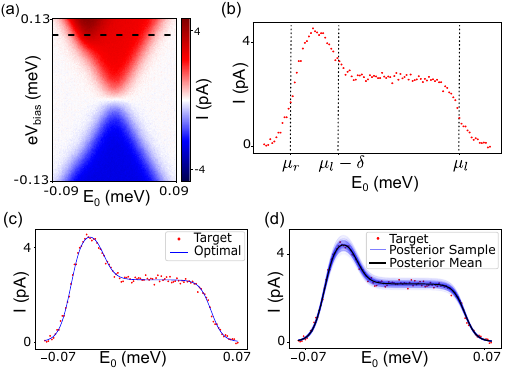}

\caption{An example of our approach applied to a single quantum dot with an orbital excited state. (a) A Coulomb diamond from which the current trace in (b) is taken, indicated by the black dashed line. (b) The example ground truth simulated current data with $100\ \mathrm{fA}$ Gaussian noise used to fit parameters, with the bias set to $0.109\ \mathrm{mV}$. Nominal locations of the points on the $E_0$ axis defined by the non-differentiable parameters $[\mu_l,\mu_r,\delta]$ are indicated by vertical dashed lines. (c) The results of gradient descent parameter optimisation on $[\Gamma_\mathrm{L},\Gamma_\mathrm{R}, T]$ after the non-differentiable parameters have been fitted. (d) The HMC posterior samples of the current values generated using 500 posterior samples of the parameters $[\Gamma_\mathrm{L},\Gamma_\mathrm{R}, T]$. Ground truth and optimal parameter values, as well as posterior means with standard deviation errors for relevant parameters, are shown in Table~\ref{tab:sqd_parameters}.}
\label{fig:sqd_fitting}
\end{figure}

To model the steady state current flowing through the SQD at a given source-drain bias $V_\mathrm{bias}$ we solve a master equation of the form,
\begin{equation}
\label{eqn:QME_sqd_excited}
    \dot{\rho} = -i[H_0,\rho] + \mathcal{L}_\mathrm{leads}\rho,
\end{equation}
where $\rho$ is the density matrix of the SQD system and $\mathcal{L}_\mathrm{leads}$ is the dissipator associated with the leads. In the weak-coupling regime $\mathcal{L}_\mathrm{leads}$ takes the form
\begin{equation}
    \begin{split}
    \mathcal{L}_\mathrm{leads} = \sum_{j=G,E} & \Big\{\big(W_{sj} + W_{dj} \big) \mathcal{D}[\vert j \rangle \langle 0 \vert]\ + \\ &
    \big( \overline{W}_{dj} + \overline{W}_{sj} \big) \mathcal{D}[\vert 0 \rangle \langle j \vert] \Big\}.
    \end{split}
\end{equation}
The tunnel rates from a lead onto the dot are given by $W_{\mathrm{L(R)}j}=\Gamma_\mathrm{L(R)}f_\mathrm{L(R)}(E_j)$ and from the dot onto the lead $\overline{W}_{\mathrm{L(R)}j}=\Gamma_\mathrm{L(R)}\big[1-f_\mathrm{L(R)}(\epsilon_j)\big]$, where $\Gamma_\mathrm{L(R)}=\pi g_\mathrm{L(R)}\vert t_{l(r)}\vert^2$ with $g_\mathrm{L(R)}$ the constant density of states in the leads in the wide band approximation, and $f_\mathrm{L(R)}(\epsilon) = [e^{(\epsilon - \mu_{l(r)})/k_B T} + 1]^{-1}$ is the Fermi-Dirac distribution for the left(right) lead.

With $e$ denoting the electronic charge and $\rho_\mathrm{ss}$ the steady state solution of Eq. (\ref{eqn:QME_sqd_excited}), the steady state current flowing though the SQD is
\begin{equation}
    I = e \sum_{j=E,G} \overline{W}_{j\mathrm{R}} \mathrm{Tr}(\rho_\mathrm{ss} \vert j \rangle\langle j \vert ) - W_{j\mathrm{R}} \mathrm{Tr}(\rho_\mathrm{ss} \vert 0 \rangle\langle 0 \vert ).
\end{equation}

\subsection*{Parameter Estimation}

We consider the model presented above as a case study to assess the performance of each part of the parameter estimation process. This model has a single controlled parameter $\boldsymbol{\theta}^\mathrm{c} = [E_0]$ which is swept through the bias window. The variable parameters $\boldsymbol{\theta}^\mathrm{v} = [\mu_l, \mu_r, \delta, \Gamma_\mathrm{L}, \Gamma_\mathrm{R}, T]$ can be separated into differentiable parameters $[\Gamma_\mathrm{L}, \Gamma_\mathrm{R}, T]$ and non-differentiable parameters $[\mu_l, \mu_r, \delta]$. The left and right chemical potentials are non-differentiable as they define the energy scale of the data before the model is computed, and the orbital excited state energy splitting $\delta$ has zero gradient for large areas of the parameter space and is thus optimised as a non-differentiable parameter. 

\begin{table}[ht]
    \centering
    \begin{tabular}{|c|c|c|c|c|}
        \hline
        Parameter & Unit & Ground Truth & Optimal & Posterior Mean $\pm$ Std. \\
        \hline
        $\mu_l$ & pixels & 15.4 & 15.3 & - \\
        $\mu_r$ & pixels & 96.6 & 96.4 & - \\
        $\delta$ & meV & 0.084 & 0.085 & - \\
        $\Gamma_\mathrm{L}$ & MHz & 18.1 & 18.0 & 18.4 $\pm$ 0.8 \\
        $\Gamma_\mathrm{R}$ & MHz & 183.1 & 180.6 & 171.3 $\pm$ 4.3 \\
        $T$ & mK & 55.9 & 53.4 & 59.4 $\pm$ 1.8\\
        \hline
    \end{tabular}
    \caption{Ground truth, optimal, and posterior means with standard deviation errors for the example shown in Figure~\ref{fig:sqd_fitting}. The non-differentiable parameters are not part of the HMC sampling and therefore do not have posterior means and uncertainties.}
    \label{tab:sqd_parameters}
\end{table}

\begin{figure*}[ht]

    \centering
    \includegraphics[width=0.96\textwidth]{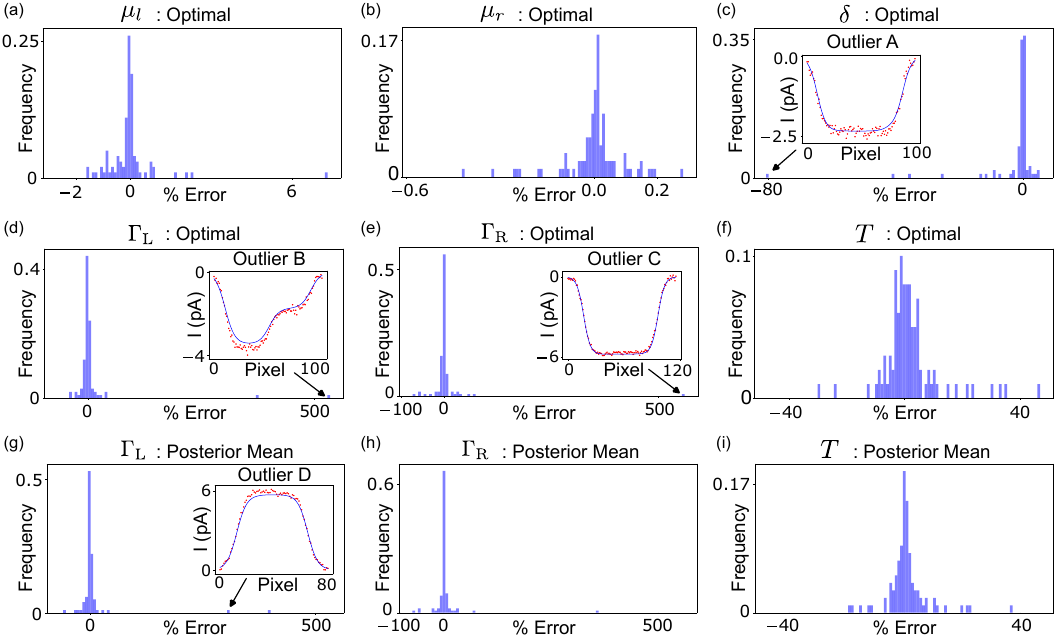}

\caption{Distributions of percentage errors for parameter estimation over a test set of 100 randomly generated current traces for a single quantum dot with an orbital excited state. Distributions displayed are: (a)-(c) optimal values of non-differentiable parameters $[\mu_l,\mu_r,\delta]$, (d)-(f) optimal values of differentiable parameters $[\Gamma_\mathrm{L},\Gamma_\mathrm{R}, T]$, and (g)-(i) posterior means for the differentiable parameters. All distributions are peaked at zero error, with a small number of outliers. Insets show the target data (red points) and optimal fit (blue line) for selected outliers. Summary statistics of these distributions can be found in Appendix~\ref{app:errors_and_posteriors}.}
\label{fig:sqd_errors}
\end{figure*}

An example of our parameter estimation algorithm is shown in Figure~\ref{fig:sqd_fitting} with parameter estimates shown in Table~\ref{tab:sqd_parameters}. In this example, the largest absolute percentage error in optimised non-differentiable values is 1.2\%, and in optimal differentiable values is 4.5\%. Posterior means and standard deviations drift slightly from the optimal parameter values, but retain a good fit to the data. In Figure~\ref{fig:sqd_errors} we consider the distributions of percentage errors for each parameter for a set of 100 iterations of our parameter estimation algorithm on randomly generated data. The distributions are all peaked at zero percentage error, indicating that our parameter estimation method is generally effective. There are a small number of outliers with large errors, but upon further inspection we find that most of these outliers have no discernible second peak in the current profile (see insets of Figure~\ref{fig:sqd_errors}, specifically Outliers A, C, and D). This means the location of the excited state cannot be determined at the first optimisation step, which impacts the estimation of further parameters. The outliers have a lower signal to noise ratio than typical samples which may further impact the parameter estimation, as evident in Outlier B which does have an excited state peak. 

From Figure~\ref{fig:sqd_errors}, we observe that parameter estimates using the posterior mean from HMC samples has a more narrow distribution around zero percentage error, and is therefore generally better than the gradient descent optimal values. The additional information contained in the posterior distribution such as spread relative to the mean, multi-modal structures, and correlations between parameters provide further insight into results of our approach. In Appendix~\ref{app:errors_and_posteriors}, we use the additional information provided by posterior distributions to filter results to remove uncertain estimates and ensure a higher accuracy in accepted results.

\section{Case Study: Double Quantum Dot}
\label{sec:case_study_double}

\subsection*{Model}

The second case study considers a double quantum dot (DQD) coupled to left and right leads and a phonon bath, as shown in Figure~\ref{fig:algorithm_sketch}(c). The states describing the DQD are $\vert 0 \rangle$, $\vert L \rangle$, and $\vert R \rangle$ where $\vert 0 \rangle$ has $(M,N)$ charges on the dot, $\vert L \rangle$ has $(M+1,N)$ charges, and $\vert R \rangle$ has $(M,N+1)$ charges. The Hamiltonian describing the coupling between these charge states, the leads, and the phonon bath is given by 
\begin{align}
    H &= H_0 + H_l + H_p + H_{el} + H_{ep}, \\
    H_0 &= \frac{\epsilon}{2}\sigma_z + t_c \sigma_x, \\
    H_l &= \sum_k (\epsilon_k - \mu_l)d_{kl}^\dag d_{kl} + (\epsilon_k - \mu_r)d_{kr}^\dag d_{kr}, \\
    H_p &= \sum_\mathbf{k} \omega_\mathbf{k} a_\mathbf{k}^\dag a_\mathbf{k}, \\
    H_{el} &= \sum_k (t_{l} d_{kl}^\dag c_l + t_{r}d_{kr}^\dag c_r + \mathrm{H.c.}), \\
    H_{ep} &= \sigma_z \sum_\mathbf{k} \lambda_\mathbf{k} (a_\mathbf{k}^\dag +a_\mathbf{k}),
\end{align}
where $\sigma_\mu$ are the Pauli operators in the subspace $\vert L \rangle$ and $\vert R \rangle$, $\epsilon=E_\mathrm{L}-E_\mathrm{R}$ is the detuning between the dot energy levels, and $t_c$ is the interdot tunnel rate. $H_l$ describes the leads with $\mu_{l(r)}$ being the chemical potential of the left(right) leads, and $d_{kl(r)}^\dag(d_{kl(r)})$ are the fermionic creation(annihilation) operators for the left(right) lead for the mode of energy $\epsilon_k$. $H_{el}$ describes the coupling between the dots and the leads where $c_{l(r)}$ is the fermionic annihilation operator acting on the left(right) dot, and $t_{l(r)}$ are the tunnel couplings between the leads and the dots. The phonon bath is described by $H_p$ where $a_\mathbf{k}^\dag(a_\mathbf{k})$ are the creation(annihilation) operators for the phonon mode of angular frequency $\omega_\mathbf{k}$. The electron-phonon interaction is described by $H_{ep}$ where $\lambda_\mathbf{k}$ is the coupling constant between the dots and the phonons.

\begin{figure}[ht]

    \centering
    \includegraphics[width=0.49\textwidth]{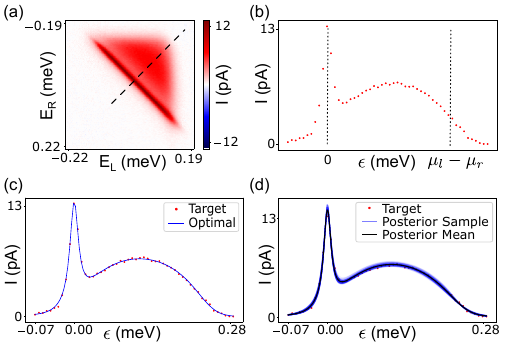}

\caption{An example of our approach applied to a double quantum dot coupled to a phonon bath. (a) A bias triangle from which the current trace in (b) is taken, indicated by the black dashed line. (b) The example ground truth simulated current data with $100\ \mathrm{fA}$ Gaussian noise used to fit parameters, with the bias set to $0.241\ \mathrm{mV}$. Nominal locations of the points on the $E_0$ axis defined by the non-differentiable parameters $[\mu_l,\mu_r]$ are indicated by vertical dashed lines. (c) The results of gradient descent parameter optimisation on $[\Gamma_\mathrm{L},\Gamma_\mathrm{R}, t_c, J_0, T]$ after the non-differentiable parameters have been fitted. (d) The HMC posterior samples of the current values generated using 500 posterior samples of the parameters $[\Gamma_\mathrm{L},\Gamma_\mathrm{R}, t_c, J_0, T]$. This example considers a 3-dimensional piezoelectric phonon spectral density with fixed parameters are $c_s=3000\ \mathrm{ms}^{-1}$, $a=20\ \mathrm{nm}$, and $d=100\ \mathrm{nm}$. Ground truth and optimal parameter values, as well as posterior means with standard deviation errors for relevant parameters, are shown in Table~\ref{tab:dqd_parameters}.}
\label{fig:dqd_fitting}
\end{figure}

To model the steady state current flowing through the DQD at a given source-drain bias $V_\mathrm{bias}=\mu_l-\mu_r$ we solve a master equation of the form
\begin{equation}
\label{eqn:QME_dqd}
    \dot{\rho} = -i[H_0,\rho] + \mathcal{L}_\mathrm{leads}\rho + \mathcal{L}_\mathrm{phonons}\rho,
\end{equation}
where $\rho$ is the density matrix of the DQD system, and $\mathcal{L}_\mathrm{leads}$ and $\mathcal{L}_\mathrm{phonons}$ are the dissipators associated with the leads and phonon bath, respectively. In the weak-coupling regime $\mathcal{L}_\mathrm{leads}$ takes the form
\begin{equation}
    \begin{split}
        \mathcal{L}_\mathrm{leads} = & W_\mathrm{L}\mathcal{D}[\vert L \rangle \langle 0 \vert] + \overline{W}_\mathrm{L}\mathcal{D}[\vert 0 \rangle \langle L \vert] +  \\ & W_\mathrm{R}\mathcal{D}[\vert R \rangle \langle 0 \vert] + \overline{W}_\mathrm{R}\mathcal{D}[\vert 0 \rangle \langle R \vert],
    \end{split}
\end{equation}
where the tunnel rates from a lead onto its neighbouring dot follow equivalent definitions as for the previous case study.

The phonon dissipator is most conveniently expressed in terms of the eigenstates of $H_0$, which for positive detuning are
\begin{align}
\label{eqn:eigenstates}
    \vert + \rangle &= \cos(\sfrac{\theta}{2}) \vert L \rangle - \sin(\sfrac{\theta}{2}) \vert R \rangle,\\
    \vert - \rangle &= \sin(\sfrac{\theta}{2}) \vert L \rangle + \cos(\sfrac{\theta}{2}) \vert R \rangle,
\end{align}
with energy $\hbar\omega_\pm\! =\! \pm\sqrt{\epsilon^2/4+t_c^2}$. The energy splitting between the eigenstates is $\hbar\omega_p\! =\! \hbar(\omega_+ - \omega_-)$, and $\theta = \arctan(2t_c/\epsilon)$. The weak-coupling phonon dissipator \cite{stace2005population} is then,
\begin{equation}
\begin{split}
    \mathcal{L}_\mathrm{phonons} = \gamma(\omega_p) & \big( [1+n(\omega_p)] \mathcal{D}[\vert - \rangle\langle + \vert] \\ & + n(\omega_p) \mathcal{D}[\vert + \rangle\langle - \vert] \big),
\end{split}
\end{equation}
where $\gamma(\omega)\! =\! \sin^2{\theta}J(\omega)$ is the rate at which the phonon bath induces interdot charge transitions in the DQD, dictated by the phonon spectral density 
\begin{equation}
\label{eqn:spectral_density_general}
    J(\omega)=2\pi\sum_\mathbf{k}\vert\lambda_\mathbf{k}\vert^2 \delta(\omega-\omega_\mathbf{k}).
\end{equation} 
The functional form of the phonon spectral density can be modelled as
\begin{equation}
\label{eqn:spectral_density_3D}
    J(\omega) = J_0 \bigg(\frac{\omega}{\omega_d}\bigg)^{n-\alpha} \bigg[ 1 - \sinc{\bigg(\frac{\omega}{\omega_d}\bigg)}\bigg]e^{-\omega^2/2\omega_a^2},
\end{equation}
where $\omega_d = \frac{2\pi c}{d}$ and $\omega_a=\frac{2\pi c}{a}$, with $d$ the centre-to-centre separation of the dots, $a$ the linear extent of the dots, and $c$ the speed of sound in the crystal, and $J_0$ is a scale factor. We also have $n$ as the dimension of the system, and $\alpha=0$ for piezoelectric coupling and $\alpha=2$ for deformation potential coupling to phonons. The complete set of spectral densities for varying dimensions and coupling mechanism is shown in Appendix~\ref{app:spectral}.

The Bose-Einstein occupation of the phonon bath $n(\omega)=[e^{\omega/k_B T}-1]^{-1}$ produces a temperature dependence on the phonon emission and absorption processes that are described by $\mathcal{D}[\vert - \rangle\langle + \vert]$ and $\mathcal{D}[\vert + \rangle\langle - \vert]$, respectively. As the DQD system is at low temperature ($k_BT\! \ll\! \omega_p$), phonon emission processes dominate.

With $e$ denoting the electronic charge, using the steady state solution of Eq. (\ref{eqn:QME_dqd}), $\rho_\mathrm{ss}$, the steady state current flowing though the DQD is
\begin{equation}
    I = e\big[\overline{W}_\mathrm{R} \mathrm{Tr}(\rho_\mathrm{ss} \vert R \rangle\langle R \vert ) - W_\mathrm{R} \mathrm{Tr}(\rho_\mathrm{ss} \vert 0 \rangle\langle 0 \vert ) \big].
\end{equation}

\subsection*{Model Selection}

We now consider the model presented above to assess the performance of our approach on a more complex system. This model has a single controlled parameter $\boldsymbol{\theta}^\mathrm{c}=[\epsilon]$ which is varied such that each dot energy level is swept through the bias window. We choose the variable parameters to be $\boldsymbol{\theta}^\mathrm{v}=[\mu_l,\mu_r,\Gamma_\mathrm{L},\Gamma_\mathrm{R},t_c,J_0,T]$ which can be separated into differentiable parameters $[\Gamma_\mathrm{L},\Gamma_\mathrm{R},t_c,J_0,T]$ and non-differentiable parameters $[\epsilon_0,\epsilon_\mathrm{bias}]$ which are the points where $\epsilon=0$ and $\epsilon=\mu_\mathrm{L}-\mu_\mathrm{R}$, respectively. The remaining parameters are set to representative fixed values of $c_s=3000\ \mathrm{ms}^{-1}$, $a=20\ \mathrm{nm}$, and $d=100\ \mathrm{nm}$. We choose these parameters to be fixed as they can all be estimated from the device design, but they could also be included in the set of differentiable parameters. The dot radius $a$ in particular has minimal impact on the fit for the energy scales probed by typical experimental measurements. The problem of estimating parameters for this model from a single current trace is expected to be underdetermined due to the number of parameters defining the system. We find fitting to data points is accurate, with an example shown in Figure~\ref{fig:dqd_fitting}. Discussion of parameter estimation for this model can be found in Appendix~\ref{app:dqd_parameters}. 

Capitalising on the accuracy of fitting to data, we use our approach to characterise other aspects of a device hosting these dots. We focus on model selection for the interaction of the double quantum dot with phonons, using our fitting process to perform model selection on the phonon spectral density. The phonon spectral density is determined by material properties and dimensionality of the semiconductor in which the dots are defined, as well as the geometry of the dots, as outlined in Appendix~\ref{app:spectral}. Accurately distinguishing between piezoelectric and deformation potential phonon coupling would yield information about the device in which the dots are formed.  Determining the dimensionality of the phonon spectral density would yield information about the confinement of phonon modes in the device and help identify any possible waveguide effects. 

\begin{figure}[ht]

    \centering
    \includegraphics[width=0.49\textwidth]{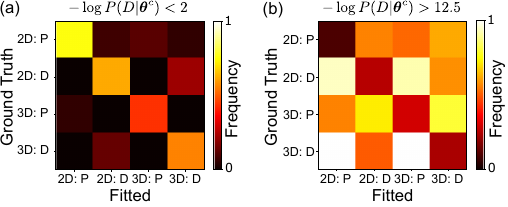}

\caption{The occurrence of fits which have a negative log likelihood of (a) less than 2, and (b) greater than 12.5 in a test set of 50 for each ground truth model. In each ground truth model a different phonon spectral density is used, with `3D: D' being 3-dimensional deformation potential, `3D: P' 3-dimensional piezoelectric, `2D: D' 2-dimensional deformation potential, and `2D: P' 2-dimensional piezoelectric. With Gaussian noise, $-\log{P(D\vert\boldsymbol{\theta}^\mathrm{c})}<2$ indicates that the average error is less than 2 noise standard deviations from the ground truth which we consider to be a good fit. Fits with $-\log{P(D\vert\boldsymbol{\theta}^\mathrm{c})}>12.5$ indicates that the average error is more than 5 noise standard deviations, representing a failure of fitting. We fit the non-differentiable parameters $[\epsilon_0,\epsilon_\mathrm{bias}]$ and the differentiable parameters $[\Gamma_\mathrm{L}, \Gamma_\mathrm{R}, t_c, J_0, T]$. Fixed parameters are $c_s=3000\mathrm{ms}^{-1}$, $a=20\mathrm{nm}$, and $d=100\mathrm{nm}$.}
\label{fig:dqd_likelihoods}
\end{figure}

In Figure~\ref{fig:dqd_likelihoods} we present the results of performing our fitting process with each phonon spectral density (2D and 3D deformation potential, 2D and 3D piezoelectric) on datasets with different ground truth spectral densities. We use the negative log likelihood, $-\log{P(D\vert\boldsymbol{\theta}^\mathrm{c})}$, as a metric for our fit. As the data has Gaussian noise, the value of the negative log likelihood indicates the number of noise standard deviations a given fit is from the true value. In Figure~\ref{fig:dqd_likelihoods}(a) we display the frequency of fits which are within two standard deviations of the true value on average. Having the largest values along the diagonal indicates that our fit is only successful when the correct model is used, providing evidence that the functional form of the spectral density adds a significant constraint on the fit. The failed fits which are on average more than five standard deviations from the true value are shown in Figure~\ref{fig:dqd_likelihoods}(b), and in this case the diagonal has the lowest values indicating that the fitting process does not fail often when using the correct model. We can also observe from this plot that the fitting process is notably worse when fitting piezoelectric spectral densities to data with deformation potential spectral densities, and also performs poorly when fitting deformation potential spectral densities to data generated using piezoelectric spectral densities in 3D. Fits which are between 2 and 5 standard deviations of the ground truth could still be classed as successful but may be more subjective, so we do not consider them useful when assessing our fitting of different spectral densities.

\begin{figure}[ht]

    \centering
    \includegraphics[width=0.48\textwidth]{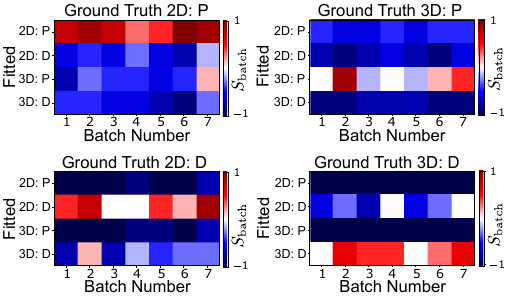}

\caption{The score of fitting to a batch of size $n_\mathrm{batch}=7$ for each phonon spectral density considered in Figure~\ref{fig:dqd_likelihoods}. The batch score $S_\mathrm{batch} = (n_\mathrm{success}-n_\mathrm{failure}) / n_\mathrm{batch}$ for seven independent batches is plotted for each fitted spectral density. A positive score indicates that the spectral density fit succeeds more than it fails within the batch, in accordance to the definitions of success and failure discussed in the text. We see that the best scores for each batch correlate with the ground truth spectral density.}
\label{fig:dqd_model_selection}
\end{figure}

We find that our approach provides useful insight into the electron-phonon coupling mechanism in a double quantum dot system, even when individual model parameters cannot be accurately determined. In order to apply this strategy to real data where the ground truth is not available, using a batch of size $n_\mathrm{batch}$ detuning traces from different bias triangles allows a decision to be made when selecting a model based on the frequencies of fit outcomes displayed in Figure~\ref{fig:dqd_likelihoods}. Defining a score function $S_\mathrm{batch}=(n_\mathrm{success}-n_\mathrm{failure})/n_\mathrm{batch}$ where $n_\mathrm{success}$ is the number of fits in a batch with $-\log{P(D\vert\boldsymbol{\theta}^\mathrm{c})}<2$ and $n_\mathrm{failure}$ is the number of fits in a batch with $-\log{P(D\vert\boldsymbol{\theta}^\mathrm{c})}>12.5$ provides reliable insight into the correct phonon coupling mechanism for simulated test data as shown in Figure~\ref{fig:dqd_model_selection} which considers the case of $n_\mathrm{batch}\! =\! 7$.

\section{Time Evolution}
\label{sec:time_evolution}

\begin{figure}[ht]

    \centering
    \includegraphics[width=0.49\textwidth]{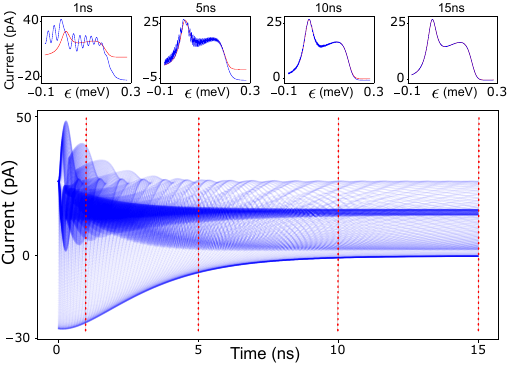}

\caption{The time-evolution of a double quantum dot system coupled to left and right leads and a phonon bath with a time-independent Hamiltonian. The upper plots display the current as a function of detuning at several points along the evolution, with the steady state current (red) shown for comparison. The evolution is computed as a batch of $n_\mathrm{b}=250$ detuning values and evolved for $15\ \mathrm{ns}$ with a step-size of $0.01\ \mathrm{ns}$. The initial state for the entire batch is the pure state, $\vert\psi(0)\rangle= (\vert 0 \rangle + \vert L \rangle + \vert R \rangle ) / \sqrt{3}$.  Parameter values used are, $V_\mathrm{bias}=0.2\ \mathrm{mV}$, $\Gamma_\mathrm{L}=\Gamma_\mathrm{R}=500\ \mathrm{MHz}$, $t_c=4\ \mathrm{GHz}$ and $T=100\mathrm{mK}$. The phonon coupling follows a 3D piezoelectric spectral density with $c_s=3000\ \mathrm{ms}^{-1}$, $a=20\ \mathrm{nm}$, and $d=50\mathrm{nm}$, and $J_0=1\ \mathrm{GHz}$.}
\label{fig:time_evolution}
\end{figure}

Having considered parameter estimation from steady-state solutions to quantum master equations using our differentiable solver, this section briefly explores time-evolution functionality and its limitations. Dynamic control of qubits is a natural motivation for studying time evolution, and thus we consider the model of a double quantum dot coupled to fermionic leads and a phonon bath outlined above as charge qubits have been implemented in such systems \cite{gorman2005charge,petersson2010quantum}. A Lindblad master equation can be recast as a set of coupled ordinary differential equations (ODEs) for time evolution solutions, which means standard ODE solvers can be used. Our solver implements a $4^\mathrm{th}$ order Runge-Kutta method \cite{butcher2016numerical} to compute time-dynamics. In Appendix~\ref{app:benchmarking}, we compare performance of our ODE solver with direct computation of the steady state as used for previous results. To verify that our solver is accurate, Figure~\ref{fig:time_evolution} shows that the steady state solution is reached after a suitable evolution time under the influence of a time-independent Hamiltonian. This figure also demonstrates batch computation of time-evolution, whereby a batch of $n_\mathrm{b}=250$ detuning values is evolved from an initial state of $\vert\psi(0)\rangle= (\vert 0 \rangle + \vert L \rangle + \vert R \rangle ) / \sqrt{3}$ over 15 ns for 1500 time steps in 0.38s using the same hardware as discussed in Appendix~\ref{app:benchmarking}.

\begin{figure}[ht]

    \centering
    \includegraphics[width=0.49\textwidth]{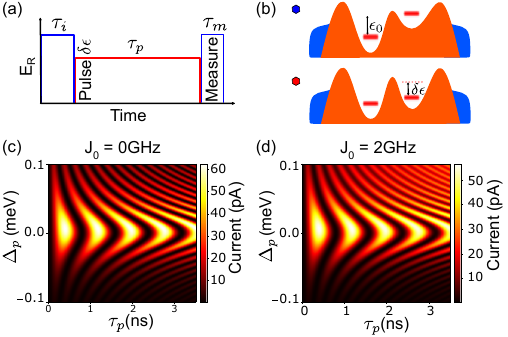}

\caption{Demonstration of a Rabi pulse sequence on a DQD charge qubit computed using our quantum master equation solver. (a) Outline of the sequence whereby the DQD is initialised to be $\vert \psi(0) \rangle = \vert L \rangle$ and after a time $\tau_i=1\ \mathrm{ns}$ a pulse is applied to the right dot to vary the detuning by $\delta\epsilon$ for a duration of $\tau_p$, and current is measured for a time $\tau_m=1\ \mathrm{ns}$. (b) Schematics of the DQD in the initialisation and measurement regime (top and the pulsed regime (bottom). Rabi chevrons without (c) and with (d) coupling to a phonon bath. We define the pulse detuning $\Delta_p=\vert\delta\epsilon\vert-\vert\epsilon_0\vert$ and the measured current is taken to be the mean current integrated during the measurement period. Parameter values used are, $\Gamma_\mathrm{L}=\Gamma_\mathrm{R}=500\ \mathrm{MHz}$, $t_c=4\ \mathrm{GHz}$ and $T=100\ \mathrm{mK}$. The phonon coupling follows a 3D piezoelectric spectral density with $c_s=3000\ \mathrm{ms}^{-1}$, $a=20\ \mathrm{nm}$, and $d=50\ \mathrm{nm}$, and $J_0=1\ \mathrm{GHz}$.}
\label{fig:rabi}
\end{figure}

In addition to evolution under a time-independent Hamiltonian, our solver can handle time-dependent Hamiltonians. We demonstrate this functionality by considering a Rabi pulse sequence on the DQD system operating as a charge qubit, with the details and results of this pulse sequence shown in Figure~\ref{fig:rabi}. We initialise the DQD such that the charge is on the left dot, with its electrochemical potential remaining below the source and drain chemical potentials for the duration of the sequence. The electrochemical potential of the right dot is pulsed by an energy $\delta\epsilon$ for a duration $\tau_p$ which defines the Rabi pulse. We then simulate the integration of current for a duration of $\tau_m$ and output the mean current during this period. Successfully reproducing the expected charge qubit dynamics under time-independent and time-dependent Hamiltonians is a demonstration that our $4^\mathrm{th}$ order Runge-Kutta solver is sufficient for the phenomena considered, however more sophisticated solvers could make computation of dynamics more efficient \cite{butcher2016numerical}. We note that while batch computation of time evolution is facilitated by our solver, providing a speed improvement over conventional solvers, gradients with respect to pulse parameters would require further work to implement a differentiable model for optimal quantum control.

\section{Conclusion}

We have presented a fast and differentiable quantum master equation solver, and demonstrated its capability to compute steady state solutions and time evolution of open quantum systems. Our solver can be applied to any system described by a Lindblad master equation, and minimal changes would permit application to a more general time-local approach involving fewer approximations. The speed of our solver comes from its capability to compute a batch of outputs in a single evaluation of a model. This speed in turn facilitates more involved algorithms requiring many evaluations, which would be impractical using standard solvers. An example of such an algorithm is the characterisation approach we develop which employs both gradient-based and gradient-free optimisation, and MCMC methods. 

Our characterisation algorithm has been applied to case studies of transport through a single quantum dot with an orbital excited state, and a double quantum dot coupled to a phonon bath. We have found our approach to be successful for dealing with parameter estimation in the case of a single quantum dot. To highlight the utility of our algorithm, we have also performed model selection to determine the electron-phonon coupling mechanism in a double quantum dot. Our approach is successful in reliably determining the ground truth spectral density. Our approach could be directly applied to experimental studies of the spectral density, such as in Ref.~\cite{hofmann2020phonon}. Decoherence from electron-phonon interactions impacts both charge and spin qubits, therefore characterising the details of this coupling mechanism is important for understanding qubit dynamics. 

Our differentiable master equation solver and its application to parameter estimation contributes to the field of characterising quantum systems using differentiable models \cite{krastanov2020unboxing, genois2021quantum, youssry2024experimental}. Our approach differs from existing methods by utilising significantly fewer data points and by dealing with a combination of differentiable and non-differentiable parameters. These attributes are key for the characterisation of more general systems where experimental data may be time-consuming or difficult to acquire. Physics-aware machine learning is a notable area in which differentiable models have been applied previously \cite{craig2024bridging,pachalieva2022physics,wright2022deep}, and our solver stands to widen the impact of such approaches, particularly in the field of quantum transport.

\section*{Code availability}

The code used in this work is available online at \url{https://doi.org/10.5281/zenodo.10783608}.

\begin{acknowledgments}
We acknowledge M. Cygorek, G.A.D. Briggs, S.C. Benjamin, and C. Sch\"{o}nenberger for useful discussions during the development of this manuscript. This work was supported by Innovate UK Grant Number 10031865, the Royal Society (URF\textbackslash R1\textbackslash 191150), the European Research Council (grant agreement 948932), EPSRC Platform Grant (EP/R029229/1), and EPSRC Grant No.~EP/T01377X/1.
\end{acknowledgments}

\appendix
\renewcommand{\theequation}{\thesection\arabic{equation}}

\section{Differentiable Solver Test Case}
\label{app:benchmarking}

To demonstrate the effectiveness of our differentiable solver, we consider the case of a single quantum dot (SQD) coupled to left and right leads. Considering the basis states $\vert 0 \rangle$ for $N$ electrons on the dot, and $\vert 1 \rangle$ for $N+1$ electrons on the dot, the total Hamiltonian is
\begin{align}
    H & = H_0 + H_l + H_{el}, \\
    H_0 & = \epsilon \vert 1 \rangle \langle 1 \vert, \\
    H_l & = \sum_k (\epsilon_k - \mu_l)d_{kl}^\dag d_{kl} + (\epsilon_k - \mu_r)d_{kr}^\dag d_{kr}, \\
    H_{el} & = \sum_k \sum_{i=l,r} t_i (d_{ki}^\dag c + d_{ki} c^\dag),
\end{align}
where $\epsilon$ is the chemical potential of the dot, $d_{kl(r)}^\dag$ and $d_{kl(r)}$ are the creation and annihilation operators for the $k^{th}$ mode in the left(right) lead, and $t_{l(r)}$ is the tunnel coupling between the dot and modes in the left(right) lead. The operators $c=\vert 0 \rangle \langle 1 \vert$ and $c^\dag=\vert 1 \rangle \langle 0 \vert$ act on the quantum dot occupation. To model the steady state current flowing through the SQD at a given source-drain bias $V_\mathrm{bias}=\mu_l-\mu_r$ we solve a master equation of the form,
\begin{equation}
\label{eqn:QME_sqd}
    \dot{\rho} = \mathcal{L}\rho = -i[H_0,\rho] + \mathcal{L}_\mathrm{leads}\rho,
\end{equation}
where $\rho$ is the density matrix of the SQD system, $\mathcal{L}$ is the Liouvillian superoperator, and $\mathcal{L}_\mathrm{leads}$ is the dissipator associated with the leads. In the weak-coupling regime $\mathcal{L}_\mathrm{leads}$ takes the form
\begin{equation}
    \mathcal{L}_\mathrm{leads}\rho = - \big( W_\mathrm{L} \mathcal{D}[c^\dag]\rho + \overline{W}_\mathrm{L} \mathcal{D}[c]\rho + W_\mathrm{R} \mathcal{D}[c^\dag]\rho + \overline{W}_\mathrm{R} \mathcal{D}[c]\rho \big),
\end{equation}
where the Lindblad superoperators act according to $\mathcal{D}[A]\rho = A\rho A^\dag - \sfrac{1}{2}\{A^\dag A,\rho\}$. The energy dependent tunnel rates from a lead onto the dot are given by $W_\mathrm{L(R)}=\Gamma_\mathrm{L(R)}f_{l(r)}(\epsilon)$ and from the dot onto the lead $\overline{W}_\mathrm{L(R)}=\Gamma_\mathrm{L(R)}\big[1-f_{l(r)}(\epsilon)\big]$, where $\epsilon$ is the electrochemical potential of the dot and $f_{l(r)}(\epsilon) = [e^{(\epsilon - \mu_{l(r)})/k_B T} + 1]^{-1}$ is the Fermi-Dirac distribution for the left(right) lead.

\begin{figure}[ht]

    \centering
    \includegraphics[width=0.45\textwidth]{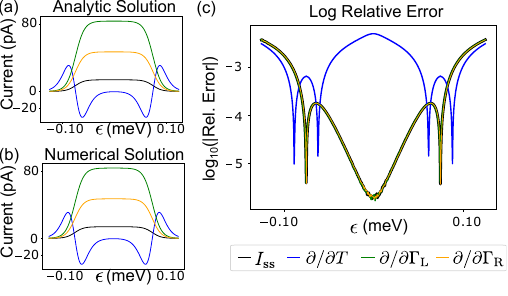}

\caption{The steady state current and its derivatives using (a) analytic solutions and (b) the differentiable solver. Parameters are set as $\Gamma_\mathrm{L}=150\mathrm{Hz}$, $\Gamma_\mathrm{R}=200\mathrm{Hz}$, and $T=100\mathrm{mK}$. A quantitative comparison of the numerical outputs and analytic solutions is shown in (c) using the logarithm of the relative absolute error. All plots share the same legend. }
\label{fig:analytic_comparison}
\end{figure}

With $e$ denoting the electronic charge and $\rho_\mathrm{ss}$ the steady state solution of Eq. (\ref{eqn:QME_sqd}), the steady state current flowing though the SQD is computed as
\begin{equation}
    I_\mathrm{ss} = e \big[ \overline{W}_\mathrm{R} \mathrm{Tr}(\rho_\mathrm{ss} \vert 1 \rangle\langle 1 \vert ) - W_\mathrm{L} \mathrm{Tr}(\rho_\mathrm{ss} \vert 0 \rangle\langle 0 \vert ) \big].
\end{equation}

This model has the benefit of a simple analytic solution for the steady state current which can be used to directly compare with the output and gradients generated by our solver. The steady state current is
\begin{equation}
    I_\mathrm{ss} = e\ \frac{W_\mathrm{L}\overline{W}_\mathrm{R} - \overline{W}_\mathrm{L} W_\mathrm{R}}{\Gamma_\mathrm{L} + \Gamma_\mathrm{R}} 
\end{equation}
with gradients with respect to $\Gamma_\mathrm{L}$, $\Gamma_\mathrm{R}$, and $T$ being
\begin{align}
    \frac{\partial I_\mathrm{ss}}{\partial \Gamma_\mathrm{L}} & = \frac{e\Gamma_\mathrm{R}^2}{(\Gamma_\mathrm{L}+\Gamma_\mathrm{R})^2} \Big[f_l(\epsilon)-f_r(\epsilon)\Big], \\
    \frac{\partial I_\mathrm{ss}}{\partial \Gamma_\mathrm{R}} & = \frac{e\Gamma_\mathrm{L}^2}{(\Gamma_\mathrm{L}+\Gamma_\mathrm{R})^2} \Big[f_l(\epsilon)-f_r(\epsilon)\Big], \\
    \frac{\partial I_\mathrm{ss}}{\partial T} & = \frac{e\Gamma_\mathrm{L}\Gamma_\mathrm{R}}{\Gamma_\mathrm{L}+\Gamma_\mathrm{R}} \Big[\frac{\partial f_l(\epsilon)}{\partial T}-\frac{\partial f_r(\epsilon)}{\partial T} \Big],
\end{align}
where
\begin{equation}
    \frac{\partial f_{l(r)}(\epsilon)}{\partial T} = \frac{\epsilon-\mu_{l(r)}}{k_B T^2}\cosh^{-2}{\Big(\frac{\epsilon-\mu_{l(r)}}{2k_BT}\Big)}.
\end{equation}

Values of $I_\mathrm{ss}$ and its derivatives computed using the differentiable solver are compared with the analytic results in Figure~\ref{fig:analytic_comparison}, displaying good agreement in relative error across the domain of dot energy $\epsilon$. We use 32 bit precision for real valued elements of the differentiable solver, with complex values defined using two 32 bit floats.

\begin{figure}[ht]

    \centering
    \includegraphics[width=0.45\textwidth]{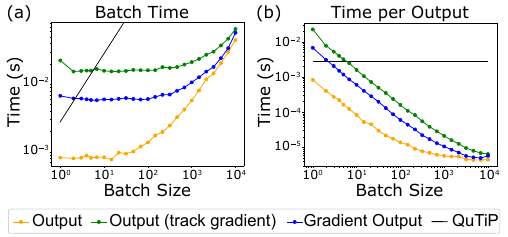}

\caption{Batch size scaling of direct computation of the steady state. The total time (a) and the time per output (b) for computations on a range of batch sizes using the differentiable master equation for the single quantum dot model described in Eq. (\ref{eqn:QME_sqd}). Each point is an average of 100 evaluation times and the time per output is simply the total time divided by the batch size. Output refers to the time taken to compute the output of a batch of input parameters. To compute a gradient by automatic differentiation, a variable must be tracked during the output computation, and once the output is available the gradient can be evaluated. Both plots share the same legend. The black lines corresponding to QuTiP execution times are extrapolated from the time taken for a single computation.}
\label{fig:time_spec}
\end{figure}

Having established that our differentiable solver returns the desired values, we now investigate the speed of the model and how computation time scales with the batch size, $n_\mathrm{b}$. The numerical results presented in Figure~\ref{fig:analytic_comparison} were performed by sending a batch of $n_\mathrm{b}=200$ parameter values with varying dot energy $\epsilon$ to our solver. We compare the evaluation of the output on the same hardware (Intel(R) Core(TM) i5-9500 CPU) with QuTiP, an established library for solving quantum master equations \cite{johansson2012qutip}. As shown in Figure~\ref{fig:time_spec}(a), our differentiable model computes batch outputs significantly faster than QuTiP (where batch computation scales linearly) across all tested batch sizes. Tracking and evaluating gradients adds an overhead to computation time which is much more noticeable at low batch sizes. Considering the computation time per output, Figure~\ref{fig:time_spec}(b) demonstrates the advantage of batched inputs with our solver where the time per output decreases to less than $10\ \mu s$. These fast computation times make our model ideal for computing a vector of outputs and evaluating the desired gradients. 

\begin{figure}[ht]

    \centering
    \includegraphics[width=0.45\textwidth]{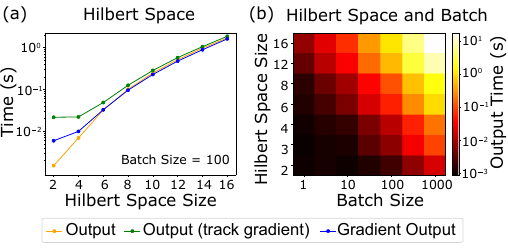}

\caption{Hilbert space scaling of the direct computation of the steady state. (a) Demonstrates the same outputs as in Figure~\ref{fig:time_spec} and how their evaluation time scales with Hilbert space size with a batch size of 100. (b) Displays the output time scaling with batch size and Hilbert space size. Each point in the average of 100 repetitions in (a) and 10 repetitions in (b). The Hilbert space dimension is increased by padding the basis vectors defining the single dot system described in Eq. (\ref{eqn:QME_sqd}).}
\label{fig:hilbert_space_time_spec}
\end{figure}

We also consider how the computation of the steady state scales with the dimension of the Hilbert space. This scaling is shown in Figure~\ref{fig:hilbert_space_time_spec}, where we find that our solver scales with Hilbert space size, $\mathcal{H}_\mathrm{dim}$, as $\sim\! \mathcal{H}_\mathrm{dim}^4$. This is as expected from the system of equations being solved in $\dot\rho(t)=0$ growing as $\mathcal{H}_\mathrm{dim}^2$. Given the time-scales shown in Figure~\ref{fig:hilbert_space_time_spec}, we expect our characterisation approach to be effective in practical times for systems of up to 6 dimensions, however larger systems could be considered if characterisation time is not a constraint. For the quantum dot systems considered in our case studies, a 6-dimensional system would equate to a chain of 5 quantum dots in the single excitation manifold.

\begin{figure}[ht]

    \centering
    \includegraphics[width=0.45\textwidth]{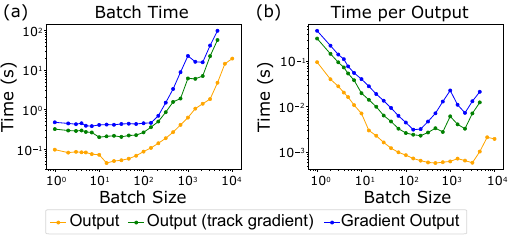}

\caption{Batch size scaling of computing the steady state using an ODE solver. The total time (a) and the time per output (b) for computations on a range of batch sizes using the differentiable master equation for the single quantum dot model described in Eq. (\ref{eqn:QME_sqd}) using an RK4 ODE solver. The ODE is solved for an initial state of $\vert 1 \rangle$ for 10 ns with 1000 time steps. Each point is an average of 100 evaluation times and the time per output is simply the total time divided by the batch size. Output refers to the time taken to compute the output of a batch of input parameters. Both plots share the same legend. The resource intensive nature of this computation leads to unexpected scaling beyond a batch size of $\sim\! 10^2$, and the last two points in each series corresponding to gradient tracking and associated output are absent as the CPU resources were exhausted for these batch sizes.}
\label{fig:time_evolution_time_spec}
\end{figure}

As our results consider steady state solutions, we compare the speed of finding the steady state using direct computation, as shown in Figure~\ref{fig:time_spec}, with a long-time integration to the steady state using an ODE solver. As discussed in the main text, we implement a $4^\mathrm{th}$ order Runge-Kutta (RK4) method in TensorFlow for time dynamics for which the scaling with batch size is shown in Figure~\ref{fig:time_evolution_time_spec}. It is clear from the comparison of batch size scaling in Figures~\ref{fig:time_spec} and \ref{fig:time_evolution_time_spec} that direct computation of the steady state is much faster and more memory efficient than using an ODE solver. More advanced ODE solvers may find some improvement over our implementation, however due to the necessarily larger number of operations in evolving an ODE it is unlikely that they would achieve equivalent performance to direct computation of the steady state. 

\section{Algorithm Scaling}
\label{app:algorithm_scaling}

\begin{figure}[ht]

    \centering
    \includegraphics[width=0.45\textwidth]{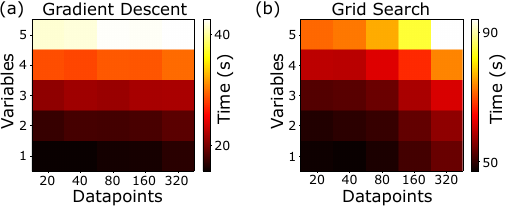}

\caption{Time taken to perform (a) gradient descent optimisation of variables and (b) the grid search over variables for the model presented in Section~\ref{sec:case_study_double}. Each point uses the same algorithm hyper-parameters as for the results in Section~\ref{sec:case_study_double} and is the average of 10 repetitions. The relevant hyper-parameters are: in (a), 2500 gradient descent steps, and in (b), 7 points in the grid search for each variable and 400 gradient descent steps to compute the loss.}
\label{fig:algorithm_scaling}
\end{figure}

We display the scaling of each constituent part of our algorithm in Figure~\ref{fig:algorithm_scaling}, namely the grid search over variables and final gradient descent optimisation. The grid search and gradient descent each take times of a similar magnitude, however the the grid search scales less favourably as can be expected from a simple grid search. Our algorithm has several hyper-parameters which can be tuned depending on requirements for speed and accuracy. The most relevant parameters for scaling are the number of points in the grid search over each variable, and the number of gradient descent steps in both the grid search and final gradient descent optimisation. The overall algorithm is then subject to the scaling of the Nelder-Mead optimiser. We also note that Bayesian optimisation is a valid alternative to a Nelder-Mead optimiser. We selected Nelder-Mead over Bayesian optimisation as the more reliable approach for our application. If one performs sequential estimates using Bayesian updates to priors using posteriors, the approach would scale linearly with the number of updates.

Gradient-free optimisation protocols, such as Nelder-Mead and Bayesian optimisation, do not scale well with number of parameters which may limit aspects of our approach to more complex systems. As discussed in Appendix~\ref{app:dqd_parameters}, we anticipate that breaking more complex systems into smaller characterisation steps with fewer parameters would circumvent this poor scaling.

\section{Errors and Posterior Filtering}
\label{app:errors_and_posteriors}

The mean and standard deviation values for the distributions displayed in Figure~\ref{fig:sqd_errors} relating to the first case study are shown in Table~\ref{tab:sqd_errors}. This table highlights that our parameter estimates are not strongly biased and have good consistency. The posterior mean errors have a lower standard deviation than optimal value errors, particularly in the case of temperature.

\begin{table}[ht]
    \centering
    \begin{tabular}{|c|c|c|}
        \hline
        &  \multicolumn{2}{|c|}{$\%$ Error (mean $\pm$ standard deviation)} \\
        \hline
        Parameter & Optimal & Posterior Mean \\
        \hline
        $\mu_l$ & $-0.01 \pm 0.64$ & - \\
        $\mu_r$ & $-0.00 \pm 0.08$ & -  \\
        $\delta$ & $-0.72 \pm 3.59$ & -  \\
        $\Gamma_\mathrm{L}$ & $-0.60 \pm 18.83$ & $-0.87 \pm 16.59$ \\
        $\Gamma_\mathrm{R}$ & $-1.61 \pm 12.62$ & $-1.76 \pm 12.24$ \\
        $T$ & $0.96 \pm 8.28$ & $-0.09 \pm 5.40$\\
        \hline
    \end{tabular}
    \caption{Summary statistics (mean and standard deviation) for the distributions displayed in Figure~\ref{fig:sqd_errors}. As the standard deviation is sensitive to outliers, we remove values greater than 3 standard deviations away from the mean before recomputing the mean and standard deviation presented here. The number of outliers identified is $1\! -\! 3\%$  for each distribution.}
    \label{tab:sqd_errors}
\end{table}

While the percentage error distributions for $\Gamma_L$ and $\Gamma_R$ have standard deviations which are not unreasonable for parameters which can vary by orders of magnitude in an experimental setting, the precision can be improved. To achieve this improvement, we capitalise on the extra information contained in the posterior distributions to reject samples with high parameter uncertainty. For each parameter posterior distribution, $P(\theta \vert D)$, we use the mean $\mu^\theta_\mathrm{post}$ and standard deviation $\sigma^\theta_\mathrm{post}$ to accept a result if all parameters fulfill the condition $\vert \sigma^\theta_\mathrm{post}/\mu^\theta_\mathrm{post}\vert \leq t$, where $t$ is a threshold value. By choosing $t\! =\! 0.05$, the error distributions are significantly narrowed such that $\Gamma_\mathrm{L}: -0.81 \pm 3.72 \%$, $\Gamma_\mathrm{R}: 1.58 \pm 6.05 \%$, and $T: 0.17 \pm 2.68 \%$, while accepting $59\%$ of results. This is a useful approach when estimation accuracy and precision are prioritised, similar to threshold choices in classifiers to optimise true-positive rates \cite{craig2024bridging}.

\begin{figure}[ht]

    \centering
    \includegraphics[width=0.45\textwidth]{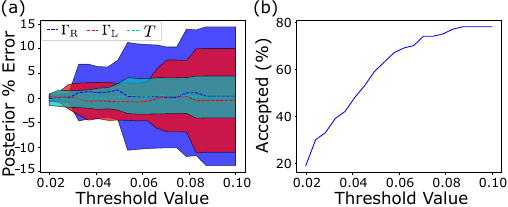}

\caption{(a) The distribution of percentage errors over a test dataset using posterior mean estimates as a function of threshold value $t$ used for filtering results. For each parameter, the dashed line is the mean and the shaded area represents the mean plus/minus the standard deviation. (b) The percentage of accepted results using the filtering condition $\vert \sigma^\theta_\mathrm{post}/\mu^\theta_\mathrm{post} \vert \leq t$ for all differentiable parameters.}
\label{fig:posterior_threshold}
\end{figure}

\section{Double Quantum Dot Parameter Estimation}
\label{app:dqd_parameters}

Table~\ref{tab:dqd_parameters} displays the parameter estimation results for the example shown in Figure~\ref{fig:dqd_fitting}. The bias is set to $0.241\ \mathrm{mV}$, and the remaining parameters are set to representative fixed values of $c_s=3000\ \mathrm{ms}^{-1}$, $a=20\ \mathrm{nm}$, and $d=100\ \mathrm{nm}$. The performance of this example is representative of our approach's parameter estimation for this model. Our approach is successful in identifying the values of $\epsilon_0$ and $\epsilon_\mathrm{bias}$ in this example, and the percentage error in these values across the case study test set is $3.48\%$ for $\epsilon_0$ and $-0.55\%$ for $\epsilon_\mathrm{bias}$. Identifying these points accurately and automatically is of benefit to experimental applications where the bias window must be known for pulsing qubits \cite{gorman2005charge, petersson2010quantum, hanson2007spins}.

The combination of an accurate fit and inaccurate parameter estimates indicate that this problem is underdetermined using the available data, as expected. To improve the performance of parameter estimation additional data would be required to constrain the fit. An example of such additional data could be a current scan along the base of the bias triangle, perpendicular to the detuning scan shown in Figure~\ref{fig:dqd_fitting}. Our Bayesian formulation then allows for posterior distributions of values for parameters from one fit to be used as priors for the subsequent fit.

\begin{table}[ht]
    \centering
    \begin{tabular}{|c|c|c|c|c|}
        \hline
        Parameter & Unit & Ground Truth & Optimal & Posterior Mean $\pm$ Std. \\
        \hline
        $\epsilon_0$ & pixels & 10.4 & 10.3 & - \\
        $\epsilon_\mathrm{bias}$ & pixels & 45.5 & 45.6 & - \\
        $\Gamma_\mathrm{L}$ & MHz & 854.1 & 127.0 & 126.2 $\pm$ 5.4 \\
        $\Gamma_\mathrm{R}$ & MHz & 166.4 & 529.5 & 540.9 $\pm$ 38.2 \\
        $t_c$ & GHz & 1.49 & 0.91 & 0.94 $\pm$ 0.04\\
        $J_0$ & GHz & 0.98 & 3.18 & 2.96 $\pm$ 0.3 \\
        $T$ & mK & 143.7 & 136.5 & 126.1 $\pm$ 24.7\\
        \hline
    \end{tabular}
    \caption{Ground truth, optimal, and posterior means with standard deviation errors for the example shown in Figure~\ref{fig:dqd_fitting}. The non-differentiable parameters are not part of the HMC sampling and therefore do not have posterior means and uncertainties.}
    \label{tab:dqd_parameters}
\end{table}

\section{Spectral Densities}
\label{app:spectral}

The following discussion provides a brief derivation of the spectral density functions considered in this article. In a spin-boson model such as that described by $H_{sb} = H_0 + H_p + H_{ep}$, and assuming the bosonic environment is in thermal equilibrium, the environment is completely described by the spectral density function $J(\omega)=2\pi\sum_\mathbf{k}\vert\lambda_\mathbf{k}\vert^2 \delta(\omega-\omega_\mathbf{k})$ as defined in (\ref{eqn:spectral_density_general}).

\begin{table*}[t]
    \centering
    {\renewcommand{\arraystretch}{2.0}%
    \begin{tabular}{|c|c|c|}
        \hline
        & \multicolumn{2}{c|}{Phonon Spectral Density: $J(\omega)$} \\
        \hline 
        Dimension &  Deformation Potential & Piezoelectric \\
        \hline 
        3D & $\frac{D^2\hbar}{2\pi^2\mu c^2 d^3}\ \Big(\frac{\omega}{\omega_d}\Big)^3\ \Big( 1- \sinc{\frac{\omega}{\omega_d}} \Big) e^{-\frac{\omega^2}{2\omega_a^2}}$ & $\frac{P^2\hbar}{2\pi^2\mu c^2 d}\ \frac{\omega}{\omega_d}\ \Big( 1- \sinc{\frac{\omega}{\omega_d}} \Big) e^{-\frac{\omega^2}{2\omega_a^2}}$ \\
        \hline 
        2D & $\frac{D^2\hbar}{4\pi^2\mu c^2 d^2}\ \Big(\frac{\omega}{\omega_d}\Big)^2\ \Big[ 1- J_0\Big(\frac{\omega}{\omega_d}\Big) \Big] e^{-\frac{\omega^2}{2\omega_a^2}}$ & $\frac{P^2\hbar}{4\pi^2\mu c^2 }\ \Big[ 1- J_0\Big(\frac{\omega}{\omega_d}\Big) \Big] e^{-\frac{\omega^2}{2\omega_a^2}}$ \\
        \hline 
        1D & $\frac{D^2\hbar}{8\pi^3\mu c^2 d}\ \frac{\omega}{\omega_d}\ \Big( 1 - \cos{\frac{\omega}{\omega_d}} \Big) e^{-\frac{\omega^2}{2\omega_a^2}}$ & $\frac{P^2\hbar d}{8\pi^3\mu c^2}\ \frac{\omega_d}{\omega}\  \Big( 1 - \cos{\frac{\omega}{\omega_d}} \Big) e^{-\frac{\omega^2}{2\omega_a^2}}$ \\
        \hline
    \end{tabular}}
    \caption{A summary of the phonon spectral density functions for deformation potential and piezoelectric interactions with a double quantum dot in varying dimensions. A similar table of spectral densities can be found in Ref. \cite{stace2005population}.}
    \label{table:spectral_densities}
\end{table*}

The interaction Hamiltonian $H_{ep}$ has a diagonal ($\sigma_z$) system operator by considering well defined quantum dots with minimal overlap of the dot wavefunctions, i.e. $\langle L \vert R \rangle\! \ll\! 1$. Using the single particle wavefunctions, the phonon coupling matrix element $\lambda_\mathbf{k}$ is defined as,
\begin{equation}
    \begin{split}
        \lambda_\mathbf{k} &= M_\mathbf{k} \big( \langle L \vert e^{i\mathbf{k}\cdot\mathbf{r}} \vert L \rangle - \langle R \vert e^{i\mathbf{k}\cdot\mathbf{r}} \vert R \rangle \big) \\
        &= M_\mathbf{k} \int\! d\mathbf{r} \big[\psi_L(\mathbf{r}) - \psi_R(\mathbf{r})\big] e^{i\mathbf{k}\cdot\mathbf{r}}
    \end{split}
\end{equation}
where $M_\mathbf{k}$ is a matrix element which depends on the material properties of the system.

Adopting the notation $\mathbf{k} = (\boldsymbol{\kappa},k_z)$ and $\mathbf{r} = (\boldsymbol{\rho},r_z)$, and assuming the dots exist in the x-y plane with identical, radially symmetric wavefunctions separated by a vector $\mathbf{d}$ such that $\psi_L(\boldsymbol{\rho},r_z)\! =\! \psi(\boldsymbol{\rho},r_z)$ $\psi_R(\boldsymbol{\rho},r_z)\! =\! \psi(\boldsymbol{\rho}-\mathbf{d},r_z)$ it follows that
\begin{equation}
\label{coupling_strength}
    \lambda_\mathbf{k} = M_\mathbf{k} \big(1-e^{i\boldsymbol{\kappa}\cdot\boldsymbol{\rho}}\big) \vert \tilde{\psi}(\boldsymbol{\kappa},k_z) \vert^2.
\end{equation}

Substituting this result into (\ref{eqn:spectral_density_general}) we can derive the form factor for the spectral density $J(\omega)$. We use a linear dispersion relation, $\omega_k\! =\! c_s\vert \mathbf{k} \vert\! =\! c_sk$ where $c_s$ is the speed of sound in the crystal, and omit the prefactor $M_\mathbf{k}$ to find the angular form factor, $F(\omega)$, based on the dot geometry,

\begin{equation}
    F(\omega) = \sum_\mathbf{k} \delta(\omega-ck) \vert\tilde{\psi}(\kappa,k_z)\vert^4 \vert 1-e^{i\kappa\cdot\mathbf{d}} \vert^2.
\end{equation}

We convert the sum over wavevectors to a 3D integral, $\sum_\mathbf{k}\! \to\! \frac{v}{(2\pi)^3} \int d\mathbf{k}$. Using the Jacobi-Anger expansion and performing the radial and azimuthal integrals, it follows that
\begin{equation}
    \begin{split}
        F(\omega) = \frac{v\omega^2}{2\pi^2c_s^3}\int_0^\pi & \vert\tilde{\psi}\Big(\frac{\omega}{c_s}\sin{\theta},\frac{\omega}{c_s}\cos{\theta}\Big) \vert^4  \times \\ &[1-J_0\Big(\frac{\omega d}{c_s}\sin{\theta}\Big)]\sin{\theta} d\theta ,
    \end{split}
\end{equation}
where $J_0(\cdot)$ is the zeroth order Bessel function of the first kind. Changing variables to $x\! =\! \cos{\theta}$ we have,
\begin{equation}
    \begin{split}
    F(\omega) = \frac{v\omega^2}{\pi^2c_s^3} \int_0^1 & \Big\vert\tilde{\psi}\Big(\frac{\omega}{c_s}\sqrt{1-x^2},\frac{\omega}{c_s}x\Big) \Big\vert^4 \times \\ &\Big[1-J_0\Big(\frac{\omega d}{c_s}\sqrt{1-x^2}\Big)\Big] dx.
    \end{split}
\end{equation}

Assuming a sharply peaked electron density simplifies the integral as $\vert\tilde{\psi}(\boldsymbol{\kappa},k_z) \vert^4\approx 1$. In the context of a double quantum dot system this is a reasonable assumption as the dots are usually confined to a narrow region of space in all directions. The finite extent of the dot wavefunctions, assumed to be Gaussian $\psi(\mathbf{r}) \propto e^{-r^2/2a^2}$, introduces a Gaussian decay from the Fourier transform of the dot wavefunction in $\ref{coupling_strength}$.

Finally, we make use of the integral $\int_0^z J_0(\sqrt{z^2-x^2})dx=\sin{z}$ to evaluate the angular form factor of the spectral density,
\begin{equation}
    F(\omega) = \frac{v}{\pi^2cd^2} \bigg(\frac{\omega}{\omega_d}\bigg)^2 \bigg( 1- \sinc{\frac{\omega}{\omega_d}} \bigg) e^{-\omega^2/2\omega_a^2},
\end{equation}
where $\omega_d = c_s/d$ and $\omega_a=c_s/a$. A similar derivation can be performed in 2D and 1D by reducing the dimensionality of the angular integral.

The prefactor $M_\mathbf{k}$ depends on the type of phonon interaction \cite{mahan2000many}. The total interaction matrix element is the sum of deformation potential and piezoelectric interaction,
\begin{equation}
    M_\mathbf{k} = \bigg( \frac{1}{2M\omega_\mathbf{k}} \bigg)^{-1/2} (D\vert \mathbf{k} \vert + iP),
\end{equation}
where M is the average mass of the unit cell, and $D$ and $P$ are the deformation potential and piezoelectric constants. Being out of phase, the contributions do not interfere to second order. The form of the respective spectral densities in (\ref{eqn:spectral_density_3D}) follows directly from these results. A summary of the functional forms of the spectral densities for deformation potential and piezoelectric phonon coupling with a double quantum dot in varying dimensions is shown in Table~\ref{table:spectral_densities}.

\bibliographystyle{naturemag}

\end{document}